# Virtual Parity-Time Symmetry

Huanan Li[1], Ahmed Mekawy[1,2], Alex Krasnok[1], and Andrea Alù[1,2,3,*]

[1]*Photonics Initiative, Advanced Science Research Center, City University of New York,
New York, New York 10031, USA*

[2]*Department of Electrical Engineering, City College of The City University of New York,
New York, New York 10031, USA*

[3]*Physics Program, Graduate Center, City University of New York,
New York, New York 10016, USA*

*Parity-time (PT) symmetry has been opening exciting opportunities in optics, yet the required careful balance of loss and gain has been hindering its practical implementations. Here, we propose a gain-free route to PT-symmetry based on non-monochromatic excitations that mimic loss and gain in passive systems. Based on the concept of virtual absorption, extended here to implement also virtual gain, we induce PT-symmetry and its landmark effects, such as broken phase transitions, anisotropic transmission resonances and laser-absorber pairs, in a fully passive, hence inherently stable, system. These results open a path to establish PT-symmetric phenomena in purely passive platforms.*

Non-Hermitian wave physics has been gaining increased attention since the discovery and demonstration of its intriguing wave phenomena, such as single-mode lasers [1]-[2], unidirectional invisibility [3]-[5], PT-symmetric laser-absorber pairs [6]-[8], anisotropic transmission resonances (ATR) [9]-[10], broken phase regimes and exceptional points (EPs) [11]-[15]. These novel phenomena have been discovered within the context of parity-time (PT) symmetry [16], a special type of symmetry that satisfies inversion upon space and time, stemming from pioneering works in theoretical quantum physics [17]-[19]. PT-symmetry has been later fruitfully applied to classical settings [15], from photonics [4], electronics [20]-[21], plasmonics [22]-[23], acoustics [24] and metamaterials [25]-[27].

---

[*]Corresponding author: aalu@gc.cuny.edu

PT-symmetric systems in classical wave physics commonly require the presence of gain. In optics, for example, PT-symmetry implies that the complex refractive index satisfies $n(\boldsymbol{r}) = n^*(-\boldsymbol{r})$, which results in a balanced gain-loss profile $\operatorname{Im} n(-\boldsymbol{r}) = -\operatorname{Im} n(\boldsymbol{r})$. This requirement has hindered several possibilities to implement and verify these concepts in practical devices, mainly because it is challenging to implement large gain in photonics, and because active systems are inherently prone to instabilities [28]-[31]. In an attempt to overcome these issues, it has been argued that PT-symmetry can be somewhat mimicked in a lossy system with unbalanced distributions of absorbing elements, by offsetting an average amount of loss [32]-[34].

Consider for instance the case of two optical modes with the same resonance frequency $u_0$, different decay rates $\gamma_0 + \gamma$ and $\gamma_0 - \gamma$, $\gamma_0 > \gamma > 0$, and a real coupling coefficient $\kappa$. The dynamics of the coupled system are described by

$$\frac{d}{dt}|\Psi\rangle = jH_0|\Psi\rangle, \quad |\Psi\rangle = \begin{pmatrix}\psi_L\\\psi_R\end{pmatrix}, \qquad (1)$$

with effective Hamiltonian $H_0 = \begin{pmatrix} u_0 + j(\gamma_0 + \gamma) & \kappa \\ \kappa & u_0 + j(\gamma_0 - \gamma) \end{pmatrix}$, where $\psi_\alpha, \alpha = L, R$ is the amplitude of the α-mode normalized such that $|\psi_\alpha|^2$ represents its energy. A hidden PT-symmetry may be revealed by offsetting the average decay rate, i.e., after the transformation $\begin{pmatrix}\psi_L\\\psi_R\end{pmatrix} =$ $e^{-\gamma_0 t} \begin{pmatrix} \psi_L' \\ \psi_R' \end{pmatrix}$, for which the Hamiltonian becomes $H_0^{PT} = \begin{pmatrix} u_0 + j\gamma & \kappa \\ \kappa & u_0 - j\gamma \end{pmatrix}$, which respects PT-symmetry with parity operator $P = \begin{pmatrix} 0 & 1 \\ 1 & 0 \end{pmatrix}$ and time-reversal operator T being the complex conjugation. In this framework, the eigenvalue spectra $\omega_\pm^{PT} = u_0 \pm \sqrt{\kappa^2 - \gamma^2}$ sustain a broken phase regime typical of PT-symmetric systems as γ crosses the EP at $\gamma_{EP} = \kappa$. This tool has been introduced as a useful path to verify some of the properties of PT-symmetry in passive systems [35], but it implies global attenuation and it requires postprocessing to actually observe the desired scattering phenomena. More complex PT-symmetric responses, such as ATRs or laser-absorber pairs, are not available in this framework simply because of passivity and power conservation.

In a different context, our group has recently introduced the concept of *virtual absorption* [36]-[38], based on which it is possible to mimic absorption in a system without loss by exciting it with non-monochromatic waves. The idea is to excite a resonant structure with a wave oscillating at a complex frequency, aligned with a scattering zero positioned in the complex frequency plane. We showed theoretically and experimentally that engaging these complex zeros enables efficient trapping of energy during the transient excitation and, when combined with nonlinearities, can also provide a powerful tool for storage and release beyond the time-bandwidth limit [39]. Here, we extend this concept to *virtual gain*. By then pairing virtual gain and loss in a balanced way, we can realize *virtual PT-symmetry* and demonstrate broken phase transitions, ATR, and laser-absorber pairs in an inherently passive and stable material platform.

*Virtual gain* — Virtual absorption can be obtained by exciting a lossless system with a signal that grows in time [36]-[38]. In a dual way, virtual gain can be achieved by exciting a passive system with a decaying signal, oscillating at a complex frequency with positive imaginary part. As a basic example, consider a balanced transmission line (TL), characterized by uniform loss $\gamma_0 > 0$ [40]- [41], and excited by a voltage signal $v_0(z,t) = V_0^+ e^{j\omega_c t - j\beta z}$ oscillating at complex frequency $\omega_c = \omega + j\sigma$, with $0 < \sigma \ll \omega$. As the decaying signal flows through the TL, the time-average power flow $J(z,t) = \frac{|V_0^+|^2}{2Z_0} e^{-2\sigma t + 2(\sigma - \gamma_0)z/v_p}$, where $Z_0$ is the characteristic impedance and $v_p$ the phase velocity. In the monochromatic regime ($\sigma = 0$), the signal decays along the line at a rate dictated by the TL loss [Fig. **1**(a)]. However, if the decay rate $\sigma$ of our signal is equal to the uniform loss $\gamma_0$, the power flow does not depend on the position $z$ at any instant in time. Similarly, when the decay rate $\sigma$ of the excitation is larger than the uniform loss $\gamma_0$, the power flow $J(z,t)$ actually grows along the propagation direction, up to the signal precursor, mimicking gain at any instant in time.

*Virtual PT symmetry*—Next, we load the TL fed by a complex frequency signal with a pair of coupled resonators, similar to those considered in Eq. (1) [see Fig. **2**(a)]. After a transient, under suitable conditions the system reaches a quasi-stationary state in which reflected and transmitted signals all decay in time following the same complex-frequency excitation. To study its dynamics, we use temporal coupled-mode theory (CMT) [42]

$$\frac{d}{dt}|\Psi\rangle = (jH_0 - \Gamma)|\Psi\rangle + D^T |s_+\rangle,$$
$$|s_-\rangle = -|s_+\rangle + D|\Psi\rangle,$$
(2)

where the input vector $|s_+\rangle = \begin{pmatrix} L^{(i)} \\ R^{(i)} \end{pmatrix}$ is formed by power-normalized amplitudes $L^{(i)}/R^{(i)}$ for incoming waves from left/right ports, and similarly the output vector $|s_-\rangle = \begin{pmatrix} L^{(o)} \\ R^{(o)} \end{pmatrix}$ refers to the outgoing waves; the 2×2 real matrix $D$ describes the coupling between ports and the two modes, and the matrix $\Gamma = \frac{1}{2}D^+ D$ accounts for the decay into the ports. For the complex frequency excitation $|s_+\rangle = e^{-\sigma t + j\omega t}|s_+^0\rangle$ with the decay rate $\sigma = \gamma_0$, i.e., matching with the average loss $\gamma_0$ of $H_0$ in Eq. (1), the scattering matrix $S$ connecting the output in the quasi-stationary state with the input via $|s_-\rangle = S|s_+\rangle$ is given by

$$S(\omega) = -I_2 + jD \frac{1}{H_0^{PT} + j\Gamma - \omega} D^T, \tag{3}$$

where $I_2$ is the 2×2 identity matrix, and the PT-symmetric Hamiltonian $H_0^{PT}$ derived above determines the internal dynamics.

The overall temporal response is determined by the interplay between the transient process, related to the initial state $|\Psi(0)\rangle$, and the quasi-stationary process determined by the scattering matrix $S$. The entire dynamics can be revealed solving Eq. (2): the outgoing waves

$$|s_-\rangle = S|s_+\rangle + De^{jH_{eff}^0 t}D^{-1}[D|\Psi(0)\rangle - (S + I_2)|s_+^0\rangle], \tag{4}$$

where the eigenvalues of the matrix $H_{eff}^0 = H_0 + j\Gamma$ represent the decay rates of the transient process. The quasi-stationary response is governed by the first term in the right-hand side of Eq. (4), whereas the role of the transient process is described by the second term, associated with the initial state $|\Psi(0)\rangle$ of the system and the initial amplitude $|s_+^0\rangle$ of the impinging waves. In order to enable virtual PT-symmetry, we carefully control the synergy between these two processes, enabling the implementation of virtual gain and loss for the complex-frequency signals.

*Virtual phase transition* — We implement these ideas in a practically viable implementation using the electronic circuit shown in Fig. **2**(a). Our circuit is composed of two coupled parallel RLC circuits with differential conductance $G_L$ and $G_R$. Their coupling is controlled by the capacitance $C_c$ in parallel with a conductance $G_c$. The signals are fed into the circuit from left and right ports by TLs with impedance $Z_0$. Virtual PT-symmetry is achieved for $|s_+\rangle = e^{-\gamma_0 t + j\omega t}|s_+^0\rangle$ when $\omega = \omega_{PT} \sim \omega_0 = 1/\sqrt{LC}$; in this framework, the Hermiticity parameter is $G$ embedded in $G_{L/R} = \pm G + \left[C + \frac{1}{L(\omega_{PT}^2 + \gamma_0^2)}\right]\gamma_0$, and we set $G_c = \gamma_0 C_c$. Correspondingly, in the quasi-stationary state the system is described by the effective PT-symmetric circuit [see Fig. **2**(a)] consisting of balanced gain/loss with resistances $\pm R = \pm 1/G$ and renormalized inductance $\tilde{L} = L\left(1 + \frac{\gamma_0^2}{\omega_{PT}^2}\right)$ [43]. The pair of resonators is capacitively coupled via $C_c$. As a first goal, we show how the system undergoes a phase transition as the Hermiticity parameter $G$ of the system is increased.

By exciting the system around the complex frequency $\omega_c = \omega + j\gamma_0$ with $\omega \sim \omega_0$ and $\hat{\gamma}_0 \equiv \gamma_0/\omega_0 \to 0$, and under the assumption of weak coupling, i.e., assuming that the coupling strengths $\hat{\varepsilon} = \frac{1}{Z_0}\sqrt{\frac{L}{C}}$ between the transmission lines and the circuit, that $\hat{c} = C_c/C$ between the resonators are of order $O(\hat{\gamma}_0)$, and that the internal loss of the resonators are small, $\hat{\gamma}_{L/R} \equiv G_{L/R}/(2C\omega_0) \leq O(\hat{\gamma}_0)$, the temporal CMT in Eq. (2) is applicable up to first order with respect to $\hat{\gamma}_0$ [41], and we can identify the parameters in Eq. (2) as $u_0 = \omega_0\left(1 - \frac{1}{2}\hat{c}\right)$, $\kappa = \omega_0\frac{1}{2}\hat{c}$, $\gamma_0 = \frac{\omega_0}{2}(\hat{\gamma}_L + \hat{\gamma}_R)$, $\gamma = \frac{\omega_0}{2}(\hat{\gamma}_L - \hat{\gamma}_R)$, and $D = \sqrt{\omega_0\hat{\varepsilon}}I_2$. A phase transition determined by $H_0$ for the passive circuit [green line in Fig. **1**(b)] is transformed to a PT-symmetric phase transition [dashed line] determined by $H_0^{PT}$ in the quasi-stationary state as $G$ grows. To complete the mapping between the two processes, the relation between voltages $v_{L/R}(t)$ at the left/right node [Fig. **2**(a)] and the mode amplitudes $\psi_\alpha, \alpha = L, R$ in Eq. (2) is given by

$$\begin{pmatrix} v_\alpha \\ \dot{v}_\alpha \end{pmatrix} = \frac{1}{\sqrt{2C}}\begin{pmatrix} 1 & 1 \\ j\omega_0 & -j\omega_0 \end{pmatrix}\begin{pmatrix} \psi_\alpha \\ \psi_\alpha^* \end{pmatrix}. \qquad (5)$$

We can also write the incoming/outgoing voltage $v_\alpha^\pm$ at node $\alpha = L, R$ in terms of the input/output in Eq. (2) as [41]

$$v_\alpha^\pm = \sqrt{\frac{Z_0}{2}}\left[\alpha^{(i/o)} + \left(\alpha^{(i/o)}\right)^*\right]. \qquad (6)$$

*Virtual ATR* — Next, we demonstrate ATR in our virtual PT-symmetric system. In the quasi-stationary process, an ATR is obtained when the reflectance of the impinging wave from one port is zero, $|S_{11}|^2 = 0$ at $\omega = \omega_{PT} = \omega_0(1 + \hat{\delta}) \in \Re$, which yields

$$\hat{\varepsilon} = (\hat{\gamma}_L - \hat{\gamma}_R) \pm \sqrt{-4\hat{\delta}(\hat{c} + \hat{\delta})} \tag{7}$$

when we apply Eq. (3) to our system. PT-symmetry $(PT) S(\omega_{PT}) (PT) = S^{-1}(\omega_{PT})$ ensures that $|S_{12}|^2 = |S_{21}|^2 = 1$ and generally $|S_{22}|^2 \neq 0$. Indeed, from Eq. (3) we find $|S_{22}|^2 = 4(\hat{\gamma}_L - \hat{\gamma}_R)^2/\hat{c}^2$. We optimize the system parameters to minimize the transient process, so that the ATR in the quasi-stationary state can be efficiently reached. To this end, we calculate the decay rates $\gamma_{tran}$ of the transient process, given by the imaginary part of the two eigenvalues of $H_{eff}^0$. For a given value of $|S_{22}|^2$, the condition that maximizes the decay rate $\gamma_{tran}^<$ of the eigenmode that lives longer is given by $\hat{\delta} = -\frac{1}{2}\hat{c}, \hat{\gamma}_R = 0, \hat{\varepsilon} = \hat{\gamma}_L + \hat{c}$, yielding

$$\gamma_{tran}^< = \gamma_0 \left[2 + \frac{2}{|S_{22}|} - Re\sqrt{1 - \frac{4}{|S_{22}|^2}}\right], \tag{8}$$

corresponding to the fastest possible transient decay for given $|S_{22}|$. This quantity monotonically decreases with $|S_{22}|$, hence to see a larger reflection contrast between the two ports we need a larger contribution from the transient process. We numerically demonstrate a virtual ATR with $|S_{22}| = 1$, for which $\gamma_{tran}^< = 4\gamma_0$, reasonably larger than the decay rate $\gamma_0$ of the excitation. In Fig. 3, we show the results of time-domain scattering simulations in the optimized system, for $\omega_0 = 2\pi \times 50\text{MHz}$, $L = 10\text{nH}$ and $\hat{\gamma}_0 = 0.01$, yielding $Z_0 = \omega_0 L/\hat{\varepsilon} \approx 52\Omega$, where $\hat{\varepsilon} \approx 6\hat{\gamma}_0$ since we have $\hat{\varepsilon} = \hat{\gamma}_L + \hat{c}, \hat{c} = 2\hat{\gamma}_L, \hat{\gamma}_L \approx 2\hat{\gamma}_0$. The initial voltage and current across capacitor and inductor in each resonator are set to zero, i.e., the system is initially at rest. In Fig. 3(a), at $t = 0$ we send an input signal (red curve) from the left port and measure the (normalized) reflected signal $|v_L^-(t)|^2/|v_L^+(0)|^2$ at the same port (green), which rapidly decays to zero after a very short transient. In contrast, the transmitted signal $|v_{L\to R}(t)|^2/|v_L^+(0)|^2$ (blue) grows and finally decays in perfect sync with the input signal as the system reaches the quasi-stationary state, confirming full transmission. Strikingly different is the response when the same signal is sent from the right port [Fig. **3**(b)]: the reflected voltage decays first due to the same transient process, but then picks up energy, and in the quasi-stationary state follows the excitation, $|S_{22}| = 1$. As expected, the transmitted signal follows the same trend as in Fig. **2**(a) due to reciprocity. During the quasi-stationary state, at any instant in time the total time-averaged power flowing out of the system is twice the incident one in this scenario, see inset in Fig. **3**(b), mimicking gain at the input port, consistent with the operation of a PT-symmetric system at the ATR point. Remarkably, we achieve this phenomenon in a purely passive system, in which the role of virtual gain is enabled by the reactive energy stored at earlier times in the system, and the suitable excitation with complex frequencies.

*Virtual absorber-laser* — Another landmark feature of PT-symmetric systems is the realization of an absorber-laser: for a specific value of Hermitcity parameter, the eigenvalues converge to support a pole and a zero at the same frequency. Here, we show a virtual absorber-laser implemented in a fully passive circuit. We require that the quasi-stationary scattering matrix $S$ in Eq. (3) possesses a pair of eigenvalues going to zero and infinity, respectively. In PT-symmetric systems, this stringent requirement can be fulfilled when the zero and pole of the $S$ matrix coalesce at $\omega = \omega_{PT} = \omega_0\left(1 - \frac{1}{2}\hat{c}\right)$, achieved in our circuit when $\hat{\delta} = -\frac{1}{2}\hat{c}$, as $\gamma$ reaches the threshold value $\gamma_{th} = \frac{1}{2}\omega_0\sqrt{\hat{\varepsilon}^2 + \hat{c}^2}$. Indeed, when $\omega = \omega_{PT}$ and $\gamma \sim \gamma_{th}$, the eigenvalues $s_\pm$ of the $S$ matrix can be approximated by $s_+ = 1/s_-^* \approx -\frac{\sqrt{\hat{\varepsilon}^2+\hat{c}^2}}{\omega_0\hat{\varepsilon}^2}(\gamma - \gamma_{th})$.

In a conventional PT-symmetric laser absorber, lasing occurs when the Hermiticity parameter reaches the lasing threshold, and past this point the system becomes unstable. For the same system, coherent perfect absorption (CPA) can be achieved when the excitation comes from both ports with amplitudes and phases matching the eigenvector $|s^0_{+,CPA}\rangle$ corresponding to the zero eigenvalue. In our scenario, the analogue of lasing corresponds to the decay of the system into its quasi-normal mode, sustained by the energy stored in the resonators at times before the quasi-stationary state. Virtual CPA, however, requires inputs with the same decay rate as the transient of the virtual lasing mode, hence it may be difficult to observe it. For this reason, we design the system to completely suppress the transient response, and employ monochromatic waves at $\omega_{PT}$ to prepare the system with suitable initial states before $t = 0$. From Eq. (4), the initial state required to avoid the transient response is $|\Psi(0)\rangle = D^{-1}(S + I_2)|s^0_+\rangle$ [44]. We first excite the system with $|s^P_+(t)\rangle = e^{j\omega_{PT}t}|s^{P,0}_+\rangle$ for $t \leq 0$, and amplitude

$$|s^{P,0}_+\rangle = \left[\left(1 + \frac{\hat{\gamma}_0}{\hat{\varepsilon}}\right) + \frac{\hat{\gamma}_0}{\hat{\varepsilon}} S(\omega_{PT})\right]|s^0_+\rangle, \tag{9}$$

which ensures reaching the initial state $|\Psi(0)\rangle$ at $t = 0$. In this preparation stage, $|s^P_-(t)\rangle = e^{j\omega_{PT}t}|s^{P,0}_-\rangle$ with

$$|s^{P,0}_-\rangle = \left[\left(1 - \frac{\hat{\gamma}_0}{\hat{\varepsilon}}\right) S(\omega_{PT}) - \frac{\hat{\gamma}_0}{\hat{\varepsilon}}\right]|s^0_+\rangle. \tag{10}$$

In Fig. **4**, we study the operation of the virtual CPA-laser around the threshold $\gamma_{th}$, i.e., $\gamma = \gamma_{th}(1 - \hat{\delta}_\gamma)$, in order to observe the absorber-laser pair as we cross the threshold. Interestingly, in contrast to conventional PT-symmetric CPA-laser, our system, being inherently passive, allows exploring also regimes beyond the threshold $\gamma_{th}$ without incurring into instabilities. Fixing $\hat{c} = \hat{\varepsilon} = \hat{\gamma}_0$, passivity $\hat{\gamma}_{L/R} \geq 0$ determines $\hat{\delta}_\gamma \geq 1 - \sqrt{2}$, which crosses the threshold value at $\hat{\delta}_\gamma = 0$. In the figure, we choose $\omega_0 = 2\pi \times 50\text{MHz}$, $L = 10\text{nH}$, $\hat{\gamma}_0 = \hat{\delta}_\gamma = 0.01$.

In Fig. **4**(a), we excite the structure with the CPA eigenvector at the two ports. Before $t = 0$ the system is excited at the real frequency $\omega_{PT}$ (red and blue lines, respectively for left and right ports), and the output signals decay as they slowly approach steady-state. As the input signals are switched to the CPA eigenvector $|s_+^{CPA}\rangle = e^{-\gamma_0 t + j\omega_{PT} t}|s_{+,CPA}^0\rangle$, $|s_{+,CPA}^0\rangle = \left(1, \ j\left[-\hat{c}^2 + \left(\frac{2\gamma}{\omega_0} - \hat{\varepsilon}\right)^2\right]/(2\hat{c}\hat{\varepsilon})\right)^T$ at $t = 0$, the system immediately transitions to the quasi-stationary CPA state, without a transient. Indeed, the output curves (green, black) suddenly go to very small values. Consistent with Eq. (10), when $S(\omega_{PT})|s_{+,CPA}^0\rangle = 0$ and $\hat{\varepsilon} = \hat{\gamma}_0$, the reflectance at time $t = 0^-$ due to the incoming monochromatic waves approaches the stationary value $|s_-^{P,0}\rangle$, which match the incoming decaying signals at time $t = 0^+$. In the inset, we show the effect of varying $\hat{\delta}_\gamma$, i.e., detuning the threshold condition, on the overall output signals $\Theta$, i.e., the ratio of total outgoing to incoming intensity. The red dot indicates the result of the main panel at time $t/\left(\frac{2\pi}{\omega_0}\right) = 15$, in which the absorption is limited to a finite value due to parasitics in the realistic circuit simulations, whereas the black curve corresponds to ideal conditions [41].

The scenario changes drastically when we consider the input $|s_+^0\rangle = (1 \quad 0)^T$. As shown in Fig. **4**(b), in this scenario the outgoing waves (green, black) are significantly larger than the incident one (red), corresponding to virtual lasing. The corresponding $\Theta-$coefficient, indicated as a red dot in the inset, along with the predicted $\Theta$ versus $\hat{\delta}_\gamma$ for ideal conditions, confirm virtual lasing. Eq. (9) also ensures that $\hat{\varepsilon} = \hat{\gamma}_0$, $|s_+^{P,0}\rangle \approx |s_-^0\rangle \equiv S(\omega_{PT})|s_+^0\rangle$, since the latter dominates near the virtual lasing threshold, confirmed by our simulations. The reflected signal $|s_-^{P,0}\rangle$ is suppressed at $\hat{\varepsilon} = \hat{\gamma}_0$, consistent with Eq. (10).

*Conclusion*s ー In this Letter, we have shown that non-monochromatic excitations engaging complex zeros and poles of a passive system can realize virtual PT-symmetry. We have demonstrated transitions from real to broken phases, ATR and CPA-laser operations in a realistic circuit configuration without the need for active elements, ensuring passivity and stability, enabling even to explore operations beyond the lasing threshold. We believe that our results may inspire the implementation and realization of PT-symmetric and non-Hermitian physics in a variety of passive photonic, phononic and electronic systems, enabling an interesting playground for classical and quantum optical phenomena without the need of stringent requirements on gain.

*Acknowledgments* ー We thank Prof. Tsampikos Kottos for stimulating discussions. This work was supported by the Air Force Office of Scientific Research and the Simons Foundation.

**Figures**

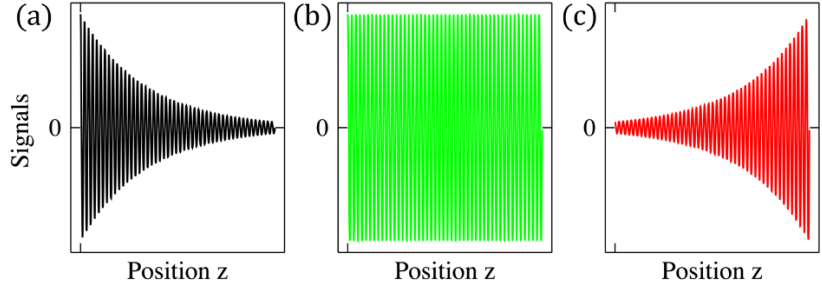

**Fig. 1.** Virtual gain: the instantaneous signals versus position $z$ in a TL with uniform loss $\gamma_0$, when the signal decay rate (a) $\sigma = 0$, (b) $\sigma = \gamma_0$, and (c) $\sigma > \gamma_0$.

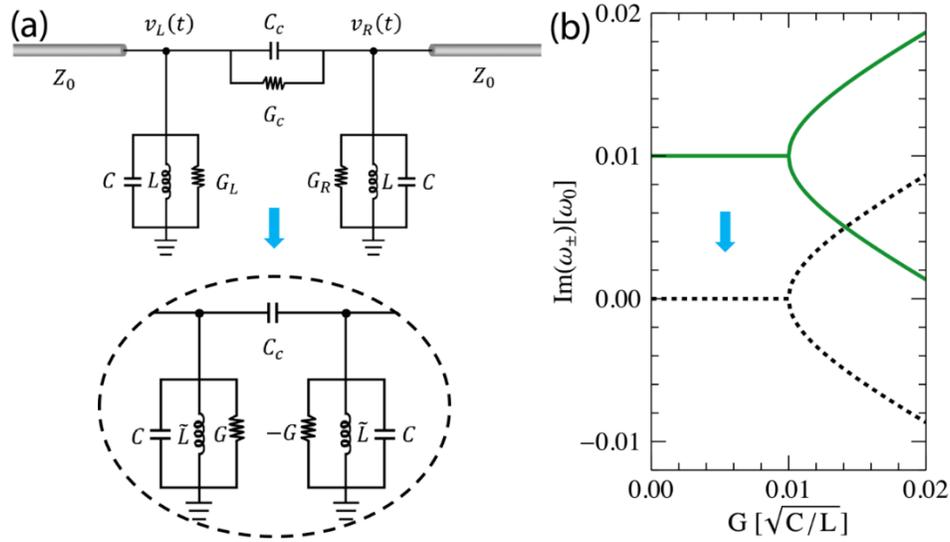

**Fig. 2.** (a) Schematic of a virtual PT-symmetric circuit. In the quasi-stationary state for suitably decaying excitation, the system is mapped onto a PT-symmetric circuit consisting of gain/loss balanced parallel $\tilde{L}RC$ resonators, with resistance $\pm R = \pm 1/G$ and renormalized inductance $\tilde{L} = L\left(1 + \frac{\gamma_0^2}{\omega_{PT}^2}\right)$, as in the dashed circle. (b) Imaginary part of the eigenvalue spectrum of the effective Hamiltonian $H_0$ of the system in (a) as a function of the Hermiticity parameter $G$. In the quasi-

stationary state, this curve is mapped onto the dashed line, corresponding to a PT-symmetric transition. Here, $\hat{\gamma}_0 = \hat{c} = 0.01$.

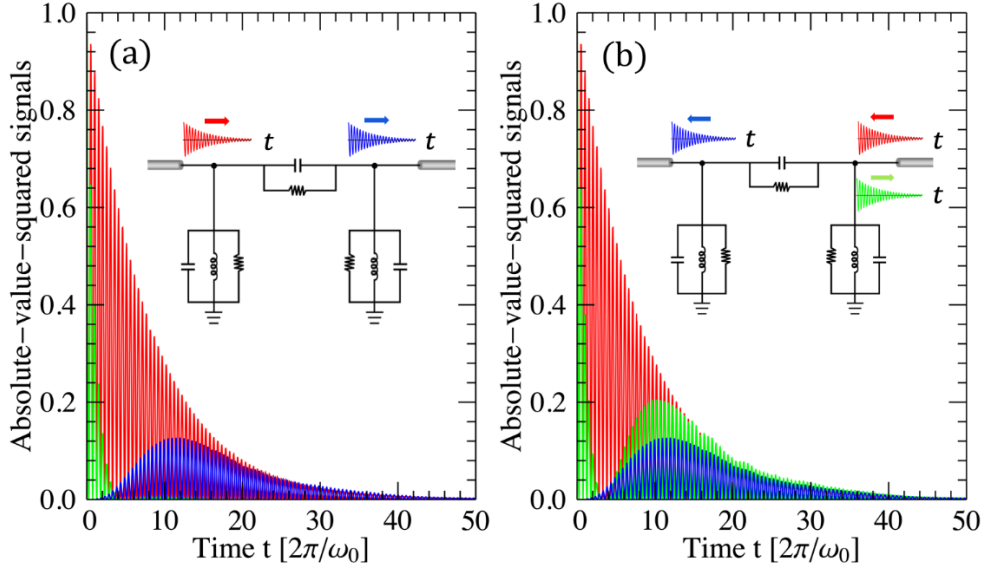

**Fig. 3.** Demonstration of virtual ATR. Time-dependent signals $v_L^+(t) = v_R^+(t)$ incident from (a) the left and (b) right port. In both panels, we plot incident (red), reflected (green), and transmitted signals (blue). In the insets, we sketch the operation in the quasi-stationary state. Here, $\omega_0 = 2\pi \times 50\text{MHz}$, $L = 10\text{nH}$ and $\hat{\gamma}_0 = 0.01$. Thus, $C = 1/(L\omega_0^2)$ and $\hat{\gamma}_R = 0$, $\hat{\varepsilon} \approx \hat{\gamma}_L + \hat{c}$, $\hat{\gamma}_L = \hat{\gamma}_0 \left[1 + \frac{1}{(1+\hat{\delta})^2 + \hat{\gamma}_0^2}\right]$, $\hat{\delta} = -\frac{1}{2}\hat{c}$ and $\hat{c} \approx 4\hat{\gamma}_0$ for ATR with $|S_{22}| = 1$. In addition, $G_1 = 2\hat{\gamma}_L\sqrt{C/L}$, $G_2 = 0$, $C_c = \hat{c}C$, $G_c = \hat{\gamma}_0\hat{c}\sqrt{C/L}$, $Z_0 = \sqrt{L/C}/\hat{\varepsilon}$, and the input signals have $\gamma_0 = \hat{\gamma}_0\omega_0$ and $\omega_{PT} = (1 + \hat{\delta})\omega_0$.

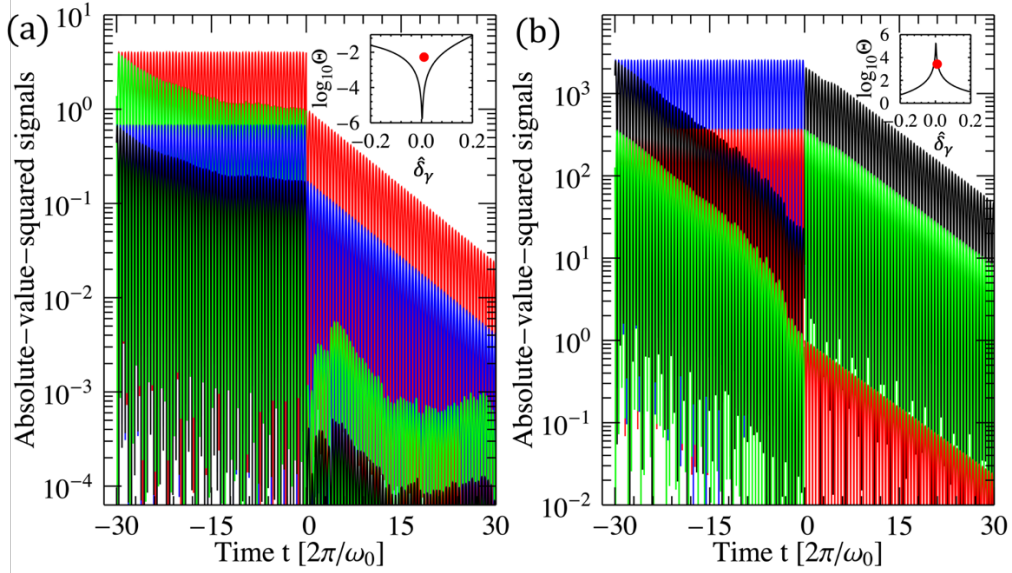

**Fig. 4.** Demonstration of absorption and amplification in a virtual CPA-laser around the threshold. The initial state $|\Psi(t=0)\rangle$ is prepared by monochromatic excitation before $t=0$, injected from left (red curve) and right ports (blue). Time-domain reflected waves at the left (green) and right (black) ports are also shown. (a) Virtual CPA after $t=0$. (b) Virtual lasing. Here, $\hat{\delta} = -\frac{1}{2}\hat{c}$, $\hat{\gamma}_{L/R} \approx \hat{\gamma}_0 \pm \frac{\gamma_{th}}{\omega_0}(1-\hat{\delta}_\gamma)$ with detuning $\hat{\delta}_\gamma = 0.01$, and $\hat{c} = \hat{\varepsilon} = \hat{\gamma}_0 = 0.01$. We choose $\omega_0 = 2\pi \times 50$MHz, L = 10nH, so that the other elements in the circuit are determined as in Fig. 3. In the insets, we show the overall output power $\Theta$ versus $\hat{\delta}_\gamma$, confirming lasing/anti-lasing after $t=0$. The result in the main panels is indicated as a red dot.

18